\documentclass{aa}  

\usepackage{graphics,graphicx}
\usepackage{xcolor}
\usepackage{enumitem}
\usepackage{caption}

\usepackage{txfonts}
\usepackage{longtable}
\usepackage[unicode=true,pdfusetitle,
 bookmarks=true,bookmarksnumbered=false,bookmarksopen=false,
 breaklinks=false,pdfborder={0 0 0},pdfborderstyle={},backref=false,colorlinks=true]
 {hyperref}
\hypersetup{
 urlcolor=blue,citecolor=blue}
\begin{document} 
  
   \title{Argon X-ray absorption in the local ISM} 
 
   \author{E. Gatuzz\inst{1}, 
           T. W. Gorczyca\inst{2}, 
           M. F. Hasoglu\inst{3},  
           J. A. Garc\'ia \inst{4} \and
           T. R. Kallman\inst{4}    
          } 

   \institute{Max-Planck-Institut f\"ur extraterrestrische Physik, Gie{\ss}enbachstra{\ss}e 1, 85748 Garching, Germany\\
              \email{egatuzz@mpe.mpg.de}
         \and
             Department of Physics, Western Michigan University, Kalamazoo, MI 49008, USA 
         \and
             Department of Computer Engineering, Hasan Kalyoncu University, 27100 Sahinbey, Gaziantep, Turkey 
         \and
             NASA Goddard Space Flight Center, Greenbelt, MD 20771, USA
             }
 
   \date{Received XXX; accepted YYY}

  \abstract  
{
We present the first comprehensive analysis of the argon K-edge absorption region (3.1-4.2 \AA) using high-resolution HETGS {\it Chandra} spectra of 33 low-mas X-ray binaries. 
Utilizing R-matrix theory, we computed new K photoabsorption cross-sections for {\rm Ar}~{\sc i}--{\rm Ar}~{\sc xvi} species.
For each X-ray source, we estimated column densities for the {\rm Ar}~{\sc i}, {\rm Ar}~{\sc ii}, {\rm Ar}~{\sc iii}, {\rm Ar}~{\sc xvi}, {\rm Ar}~{\sc xvii} and {\rm Ar}~{\sc xviii} ions, which trace the neutral, warm and hot components of the gaseous Galactic interstellar medium.  
We examined their distribution as a function of Galactic latitude, longitude, and distances to the sources. 
However, no significant correlations were discerned among distances, Galactic latitude, or longitude. 
Future X-ray observatories will allow us to benchmark the atomic data as the main resonance lines will be resolved.  

}
 
    \titlerunning{Argon X-ray ISM absorption}
    \authorrunning{Gatuzz et al.}
   \maketitle

%________________________________________________________________

\section{Introduction}\label{sec_in}
The interstellar medium (ISM) stands as a fundamental constituent in Galactic dynamics, influencing star life cycles, regulating cooling processes in molecular clouds, which enables star formation \citep{won02,big08,ler08,lad10,lil13}. 
Comprising gas and dust dispersed between stars, the ISM manifests in multiple phases characterized by distinct gas temperatures spanning from 10 to 10$^{6}$ K \citep[e.g.,][]{mck77,fal05,ton09,dra11,jen11,rup13,zhu16,sta18}. 

High-resolution X-ray spectroscopy emerges as a potent tool for dissecting  the complex environment of the ISM. 
This analytical technique requires bright X-ray sources functioning akin to beacons. 
By modeling absorption features discerned in X-ray spectra, we can study the physical properties of the intervening gas along the line-of-sight to the source. 
The advent of X-ray observatories equipped with grating spectrometers capable of high-resolution spectra has propelled ISM X-ray absorption studies to the forefront of X-ray astronomy.
These investigations encompass the study of K photoabsorption edges across various elemental species such as O, Fe, Ne, Mg, N, Si, and S \citep{pin10,pin13,cos12,gat13a,gat13b,gat14,gat15,gat16,joa16,gat18a,gat18b,gat18c,rog18,gat19,zee19,gat20,psa20,gat21,rog21,yan22,gat23,gat24}.  Furthermore, L-edge studies of Fe have also been carried out in the last decade \citep{cos12,wes19,psa23,cor24,psa24}. 

Argon holds significant importance within the ISM due to its abundance and ability to trace various physical and chemical processes. 
Notably, argon serves as a diagnostic tool in shock regions, aiding in studying shock wave interactions within supernova remnants \citep{dop18,dop19}. 
Its depletion onto dust grains influences the composition and properties of interstellar dust, thereby impacting processes such as dust grain growth \citep{are14,ama21,jon23}. 
Furthermore, argon acts as a tracer of cosmic ray ionization in the ISM, contributing to our understanding of cosmic ray propagation and their effects on the surrounding medium \citep{ogl09}. Observations of argon emission lines in star-forming regions help identify mechanisms triggering star formation \citep{lop10,str23}. 
Additionally, the abundance of argon relative to other elements, such as hydrogen and oxygen, provides valuable information for studying the metallicity of different regions within the ISM \citep{hua23}. 
In comparison with studies of oxygen and sulfur, the amount of argon depletion into dust grain is more challenging.
In their calculation of ionization correction factors for argon in giant H II regions, \citet{ama21} provide only a qualitative analysis of the depletion into dust, given that argon stellar abundances are highly uncertain.
Furthermore, although argon is mainly produced by core-collapse supernovae, Type Ia supernova may also contribute to its production \citep{san13,leu18}.
In the Galactic chemical evolution by \citet{kob20}, they found that up to $34\%$ of the argon in the solar vicinity can have Type Ia supernova origin.
Therefore, studies of argon depletion onto dust must include chemical evolution models considering both supernova yields, despite their uncertainties \citet{pal21}.

In this study, we analyzed of the Ar K-edge absorption region utilizing {\it Chandra} observations of low-mass X-ray binaries (LMXBs). 
Section~\ref{sec_xray_data} outlines the data sample and the spectral fitting procedure employed. 
Section~\ref{sec_ar_atom} delves into the computation of atomic data and the incorporation of photoabsorption cross-sections within our modeling framework.
The outcomes of our fits are discussed in Section~\ref{sec_dis}. 
Finally, Section~\ref{sec_con} provides a concise summary of our analysis.

\begin{figure} 
\centering
\includegraphics[scale=0.32]{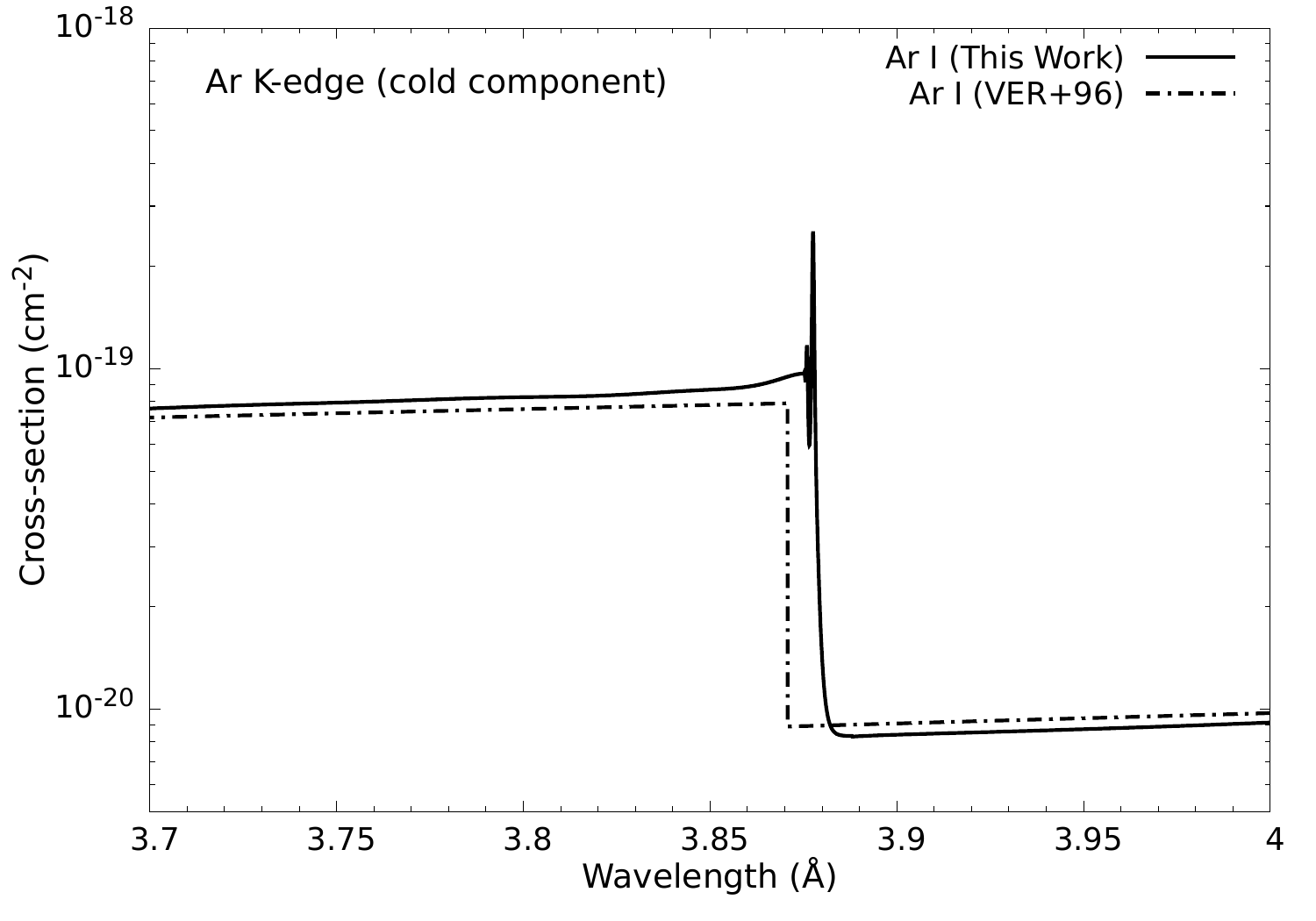}\\
\includegraphics[scale=0.32]{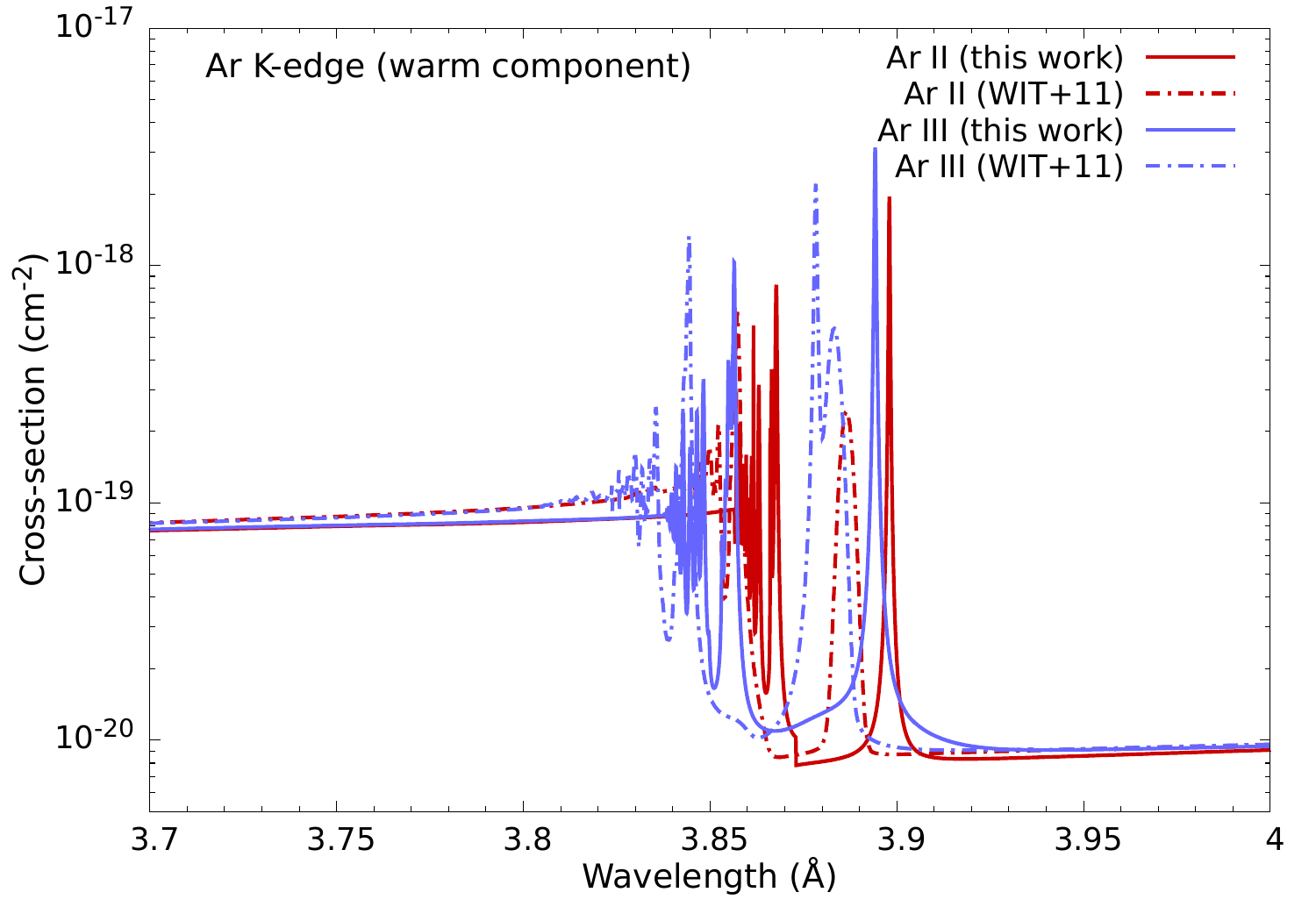}\\
\includegraphics[scale=0.32]{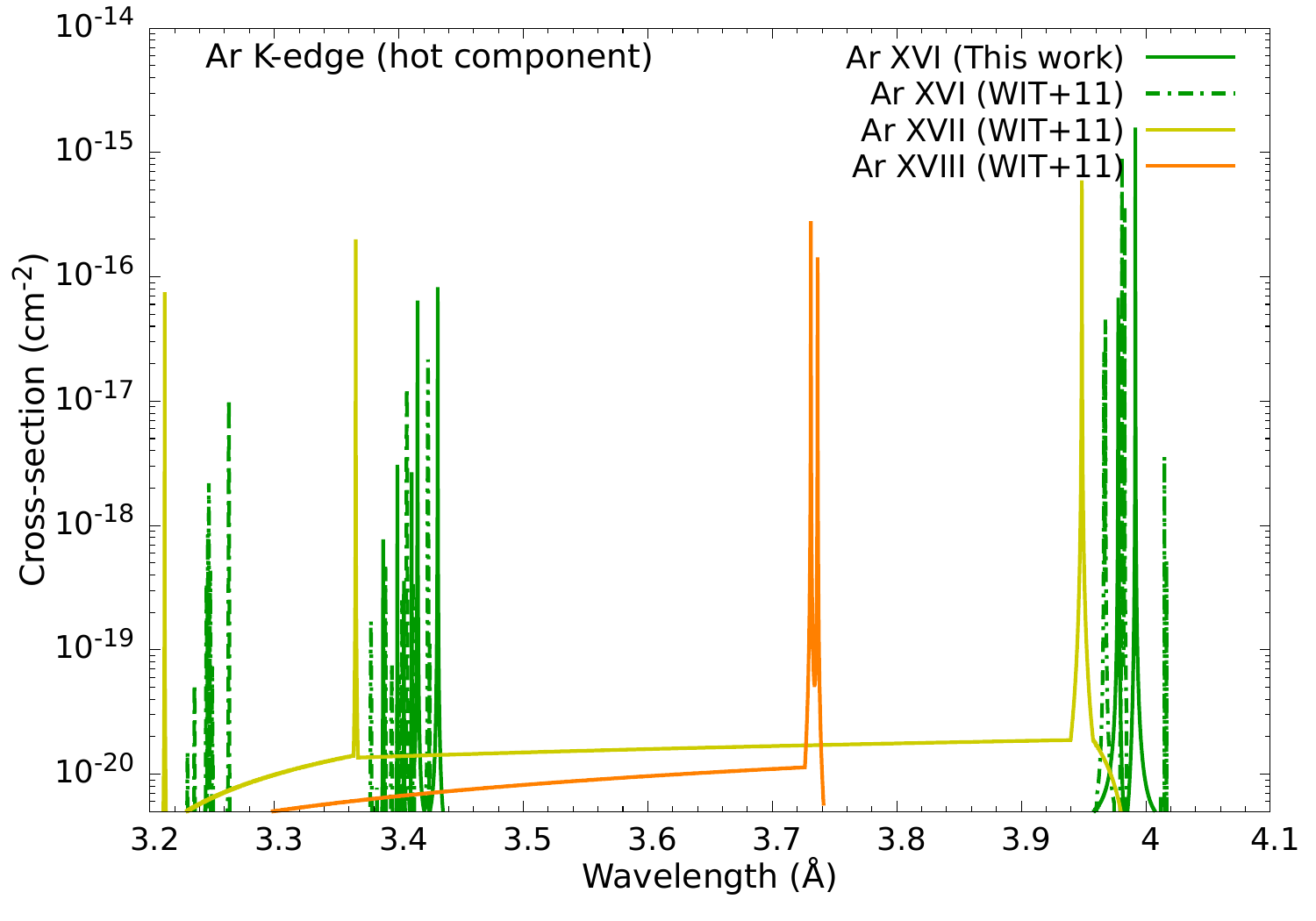} 
\caption{ Photoabsorption cross sections included in the model for {\rm Ar}~{\sc i} (top panel), {\rm Ar}~{\sc ii}, {\rm Ar}~{\sc iii} (middle panel), {\rm Ar}~{\sc xvi}, {\rm Ar}~{\sc xvii} and {\rm Ar}~{\sc xviii} (bottom panel). The plots also include {\rm Ar}~{\sc i} cross-section from \citet{ver96b} and {\rm Ar}~{\sc ii}, {\rm Ar}~{\sc iii}, {\rm Ar}~{\sc xvi} cross-sections from \citet{wit11}.  }\label{fig_ar_cross}
\end{figure} 
\section{Data reduction and analysis}\label{sec_xray_data}   
We analyzed {\it Chandra} spectra from 33 low-mass X-ray binaries (LMXBs) observed along various lines of sight. 
Our sample selection criteria required a minimum of 1000 counts in the argon edge absorption region (3.1--4.2 \AA). 
 This criterion has been applied in previous analysis \citep[e.g.,][]{gat21,gat23,gat24} and allow us to gather a diverse sample of sight-lines, without preconceived assumptions about the strength of Argon absorption.
While the column density ($N({\rm HI+H2})$) is an important factor in X-ray transmission and absorption, imposing a constraint based on $N({\rm HI+H2})$ could introduce biases by preferentially selecting sight-lines with certain characteristics.
Our approach, based on counts, include a wider range of column densities and a larger number of sources, therefore providing a more representative sample of the ISM conditions.
We deliberately avoided biasing the sample by refraining from imposing constraints based on specific line detections, such as {\rm Ar}~{\sc xvii} K$\alpha$. 

Table~\ref{tab_lmxbs} provides specifications for the selected sources, including Galactic coordinates, hydrogen column densities adopted from \citet{wil13}, and, when available, distances. 
Although the column densities from \citet{wil13} represent average values of $N({\rm HI+H2})$ along sightlines traversing the entire Galaxy, these values are appropriate for our analysis due to the narrow wavelength region covered ($\sim 1$ \AA).  Appendix \ref{sec_app} shows the observations IDs analyzed for each source.

All spectra were acquired using the High Energy Grating (HEG) in combination with the Advanced CCD Imaging Spectrometer (ACIS) onboard {\it Chandra}. 
Data reduction procedures, including background subtraction, were performed using the Chandra Interactive Analysis of Observations (CIAO\footnote{http://cxc.harvard.edu/ciao/threads/gspec.html}, version 4.15.1),  following standard CIAO procedure for point-like sources. We used the {\tt findzo} algorithm\footnote{https://space.mit.edu/cxc/analysis/findzo/} to estimate the zero-order position for those sources where zero-order data were not telemetered. Then, we used the {\tt chandra\_repro} script which reads data from the standard data distribution (e.g. primary and secondary directories) and creates a new bad pixel file together with and event file and type II PHA files, response files and auxiliary response files.   

Spectral fitting was conducted within the 3.1--4.2 \AA\ wavelength range using the {\sc xspec} package (version 12.11.1\footnote{http://heasarc.nasa.gov/xanadu/xspec/}). 
We employed a {\tt powerlaw} model to characterize the continuum, with parameters such as $\gamma$ and normalization treated independently across all observations to accommodate potential variations over different epochs. 
The goodness-of-fit assessment utilized $\chi^{2}$ statistics, complemented by the weighting method proposed by \citet{chu96}.

\section{Photoabsorption cross section calculations}\label{sec_ar_atom}
In order to fit the Ar K-edge absorption region, atomic \ion{Ar}{i}-\ion{Ar}{xvi} cross sections are computed by using R-matrix theory. Within a single-configuration perspective for the K-shell photoexcitation processes in the argon ground state, the following configurations are involved:
\begin{eqnarray}
h\nu+S(1s^22s^22p^63s^23p^6)[ ^1S ]\rightarrow 1s2s^22p^63s^23p^6np[ ^1P^o ]\nonumber \ .
\end{eqnarray}

Using these orbitals, and some additional pseudoorbitals that are optimized on the $1s$-vacancy states and account for relaxation effects, the following list of target states and computed R-matrix energies are compared to available NIST values (see Table~\ref{Ar1en}).  The $2p^{-1}$ L-edge states are autoionizing themselves and are not available from the NIST atomic database~\citep{nist}.

\begin{table}[h]
\caption{Energies (in Rydbergs) for the Ar~{\sc i} ground state and the Ar~{\sc ii} target states. }
\label{Ar1en}      
\centering                
\begin{tabular}{l l c c}  
\hline\hline              
 Ion & State & R-Matrix$^a$ & NIST$^b$  \\
Ar~{\sc i}   & $1s^22s^22p^63s^23p^6 \ ^1S$     &  -1.106894   & -1.158310 \\
\hline 
Ar~{\sc ii}  & $1s^22s^22p^63s^23p^5 \ ^2P^o$   &   0.000000   &  0.000000 \\
             & $1s^22s^22p^63s3p^6   \ ^2S$     &   1.066882   &  0.986395 \\
             & $1s^22s^22p^53s^23p^6 \ ^2P^o$   &  17.416691   &     -     \\
             & $1s^22s2p^63s^23p^6   \ ^2S$     &  22.810213   &     -     \\
             & $1s2s^22p^63s^23p^6   \ ^2S$     & 234.032658   &     -     \\
\hline\hline 
\end{tabular}
%\tablefoottext{a}{Present work}
%\tablefoottext{b}{NIST spectroscopic values \citep{NIST}}
\end{table}

Those high-energy intermediate autoionizing states can decay via two fundamentally distinct Auger pathways. The first decay route is via {\em participator} Auger decay in which the valence electron $np$ primarily takes part in the autoionization process,
\begin{eqnarray}
1s2\ell^{\, 8} 3\ell^{\, 8} np & \rightarrow & 1s^2 2\ell^{\, a} 3\ell^{\, b} +e^-\ , \ \ (a+b=15) \nonumber \ ,
\label{eqpart}
\end{eqnarray}
yielding a decay rate that scales as $1/n^3$.
The second decay pathway is {\em spectator} Auger decay in which the valance electron $np$ doesn't take part in the autoionization process,
\begin{eqnarray}
1s2\ell^{\, 8} 3\ell^{\, 8} np & \rightarrow & 1s^2 2\ell^{\, c} 3\ell^{\, d}np +e^-\ , \ \ (c+d=14) \nonumber \ ,
\label{eqspect}
\end{eqnarray}
for which the Auger rate is independent of $n$.   {\em Spectator} Auger decay becomes the dominant decay pathway at higher $n$ values in photoexiation process and becomes constant at all $n$, leading to a significant broadening of entire Rydberg series of resonances below the K-edge and an apparent K-edge significantly below the actual K-shell ionization thresholds. In the standard R-matrix implementation \citep{burke,berrington95}, participant Auger decay is explicitly taken into account by including all final \ion{Ar}{ii} target states. 

For {\em spectator} Auger, on the other hand, it is impossible to include all of the infinite $1s^2 2\ell^{\, c} 3\ell^{\, d}np +e^-\ , \ \ (c+d=14)$ decay channels implicitly to account for spectator Auger decay effects within the standard R-matrix implementation. Instead, the present calculations utilize the modified R-matrix method to account for  spectator Auger broadening effects by using an optical potential, as described by \citet{Gorczyca99,burke}. In this approach, the target energy of each closed channel is modified, within a multi-channel quantum defect theory approach, as
\begin{equation}
E  \rightarrow E -i\Gamma/2  \nonumber \ ,
\label{eqbroad}
\end{equation}
where $\Gamma$ is the $1s2s^22p^63s^23p^6$ Auger width.

The Auger widths needed to treat spectator Auger broadening effects in $1s^{-1}$ autonionizing target states are computed by using the Wigner Time Delay Method \citep{smith}. Specifically, the R-matrix method is employed on the $e^-+1s^2n\ell^{q-2}$ scattering problem, similar to photoabsorption calculations in terms of basis set and configuration lists. Auger widths are then obtained by analyzing scattering channels. The Auger width for the $1s2s^22p^63s^23p^6$ autonionizing state is determined to be $3.90\times 10^{-2}$ Ryd.  
This modified $R$-matrix method with pseudoresonance elimination \citep{pseudo} is applied to all \ion{Ar}{i}-\ion{Ar}{xiv} ions, similar to earlier benchmarking with experimental synchrotron measurements \citep{Gorczyca99,goro,gor13,gorne,hasoglu_c} and high resolution spectroscopic observations \citep{hasoglu_c,mg,gatuzz_si,gatuzz_s}.

In order to account for relaxation effects due to $1s$-vacancy, the orbital basis used in the implementation of R-matrix calculations consisted of physical orbitals and peseudoorbitals. For the present study, the number of electrons in the target states ranges from$ N =0,1, ... ,10 $, and the orbitals consisted of $1s$, $2s$, and $2p$ physical orbitals and $\overline{3s}$, $\overline{3p}$, and $\overline{3d}$ pseudoorbitals. For $ N \ge 10 $, a larger basis is needed to treat $n=3$ resonances accurately. Therefore, $1s$, $2s$, $2p$, $3s$, and $3p$ are treated as physical orbitals and nd $\overline{3d}$, $\overline{4s}$, and $\overline{4p}$ are pseudoorbitals. A Hartree-Fock method is employed to compute the physical orbitals. On the other hand, the pseudoorbitals, that are optimized to account for important orbital relaxation effects, are computed by utilizing a multi-configuration Hartree-Fock method on the configuration lists $n=3$-complex ($ N < 10 $) and $n=4$-complex ($ N \ge 10 $)  formed by single- and double-promotions from the $1s$-vacancy configuration.

\begin{figure}
          \centering  
\includegraphics[width=0.48\textwidth]{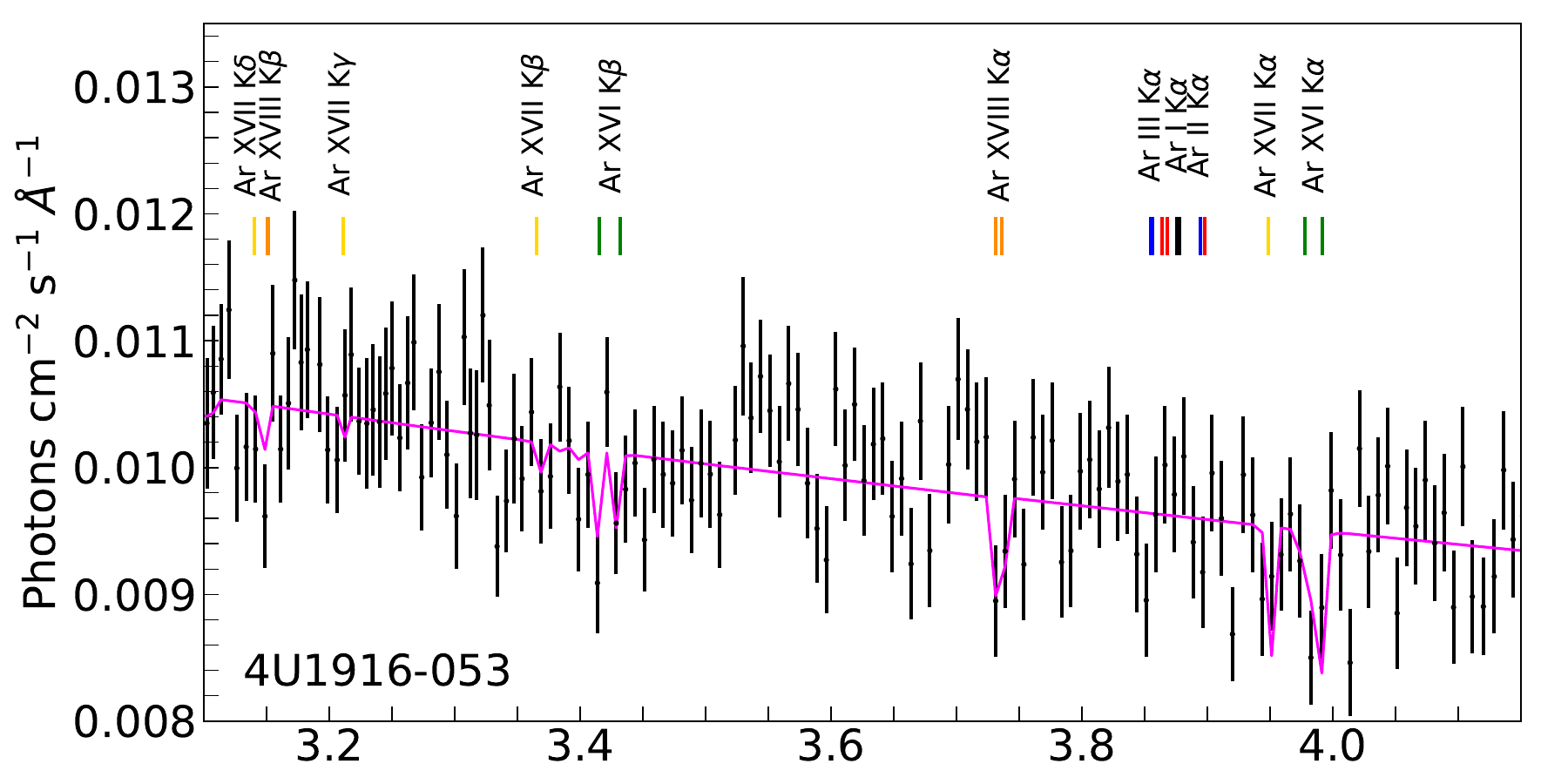} \\  
\includegraphics[width=0.48\textwidth]{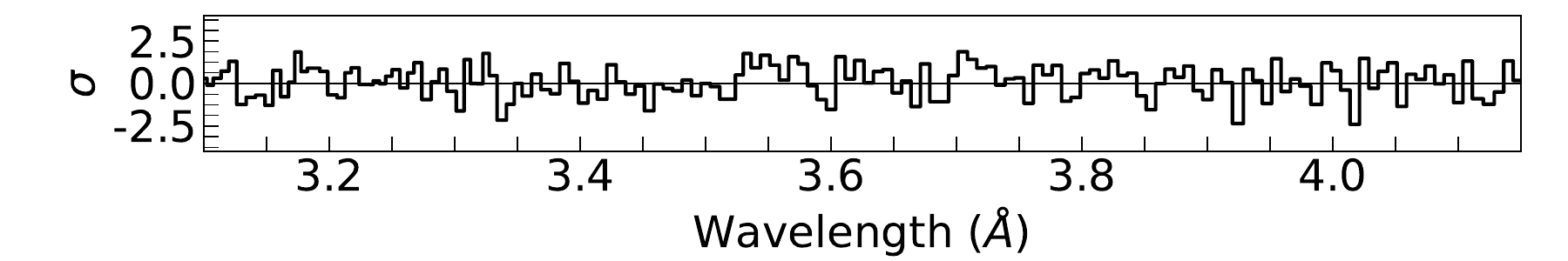}   
      \caption{
      Best-fit results in the Ar K-edge photoabsorption region for the LMXB 4U~1916-053. Black points correspond to the observation in flux units, while the red line corresponds to the best-fit model. Residuals are included in units of $(data-model)/error$. The position of the K$\alpha$ absorption lines are indicated for each ion, following the color code used in Figure~\ref{fig_ar_cross}.
      }\label{fig_fits_4U1916-053}
   \end{figure}

To characterize X-ray absorption in the ISM due to argon Ar, we utilize the photoabsorption cross-sections for {\rm Ar}~{\sc i}, {\rm Ar}~{\sc ii}, {\rm Ar}~{\sc iii}, and {\rm Ar}~{\sc xvi} K-edge, as previously described. 
Additionally, we incorporate the photoabsorption cross-sections for {\rm Ar}~{\sc xvii} and {\rm Ar}~{\sc xviii} from \citet{wit11}. 
These cross-sections are depicted in Figure~\ref{fig_ar_cross} and integrated into a modified version of the {\tt ISMabs} model \citep{gat15}\footnote{\url{https://github.com/efraingatuzz/ISMabs}}. The column densities of atomic hydrogen (H{\textsc{i}}) in {\tt ISMabs} are constrained to the values provided by \citet{wil13} for each source. 
For comparison we also included {\rm Ar}~{\sc i} cross-section from \citet{ver96b} and {\rm Ar}~{\sc ii}, {\rm Ar}~{\sc iii}, {\rm Ar}~{\sc xvi} cross-sections from \citet{wit11}. 
\citet{ver96b} cross-section does not include any resonance lines but only the K-edge. 
We note that previously reported K-absorption cross-sections of Argon ionized species computed by \citet{wit11} while utilizing a similar R-matrix approach with the inclusion of Auger broadening, important orbital relaxation effects were not included because the single-electron orbitals were obtained by using a Thomas-Fermi-Dirac statistical model potential.
Relaxation effects affect K-shell threshold estimation, as shown by the overestimation by $\sim7$~eV of the Ar~II and Ar~III K-edges..
For instance, Figure~\ref{fig_fits_4U1916-053} illustrates the best fit achieved for the LMXB 4U~1916-053. 
Due to the nominal HEG resolution of $\Delta\lambda\sim 12$ m\AA, surpassing the separation of the K$\alpha$ resonance lines, a detailed benchmarking of atomic data proves challenging.  
This limitation suggests a potential for future analyses leveraging high-resolution X-ray instrumentation (see Section~\ref{sec_sim}).

           \begin{figure}
          \centering
\includegraphics[width=0.43\textwidth]{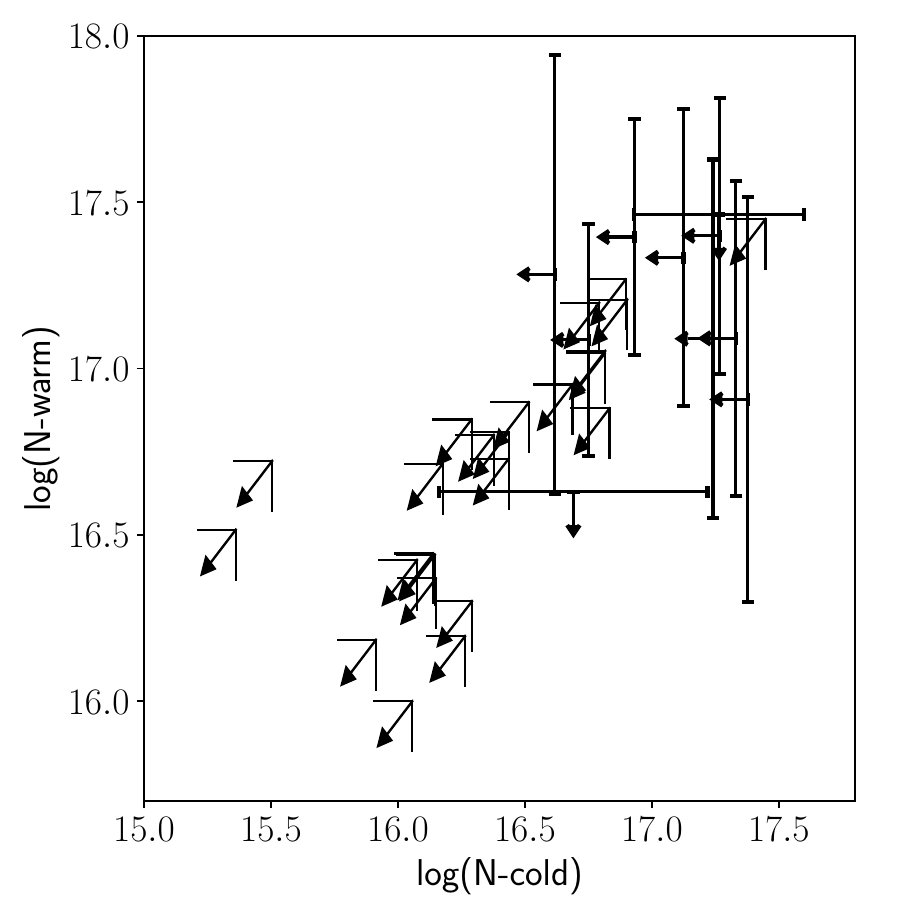}\\
\includegraphics[width=0.43\textwidth]{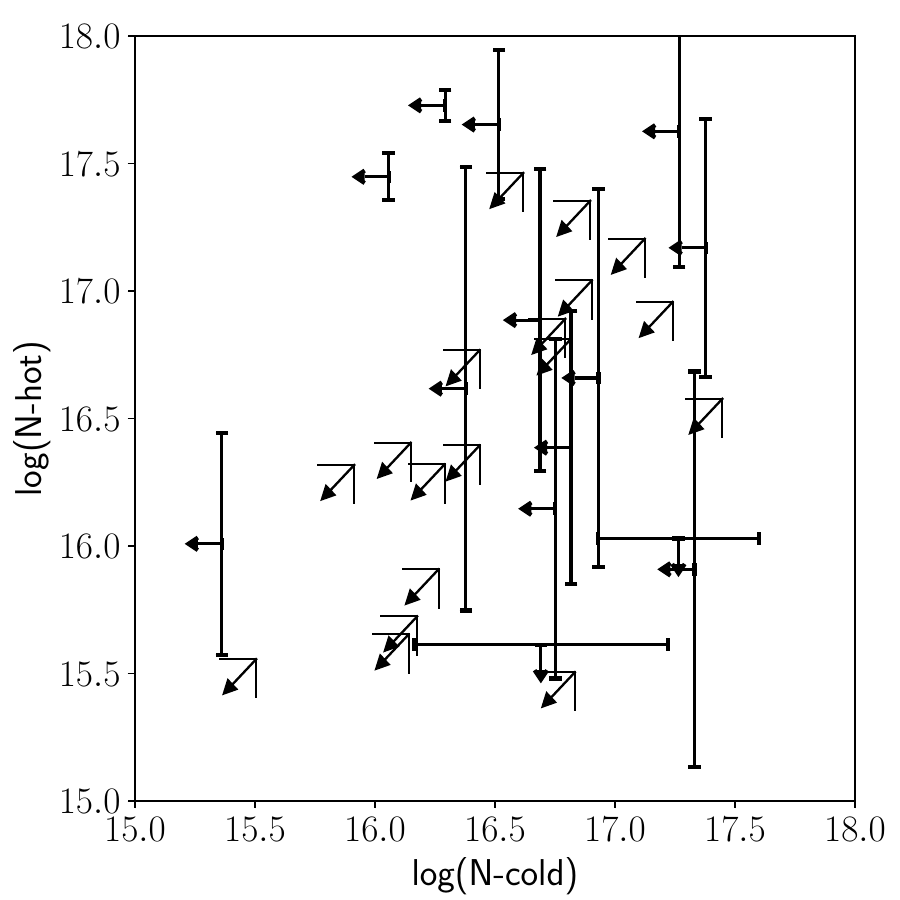}   
      \caption{
      Best fit column densities for the cold ({\rm Ar}~{\sc i}), warm ({\rm Ar}~{\sc ii}+{\rm Ar}~{\sc iii}), hot ({\rm Ar}~{\sc xvi}+{\rm Ar}~{\sc xvii}+{\rm Ar}~{\sc xviii}) ISM phases.
      }\label{fig_data_density_fractions}
   \end{figure}
\begin{table}
\caption{\label{tab_lmxbs}List of Galactic observations analyzed.}
%\tiny
\centering
\begin{tabular}{lccc}
\hline
Source   & Galactic &Distance     &$N({\rm HI+H2})$\\
 &Coordinates &(kpc)   &     \\
\hline  
4U~0614+091 &$( 200.88 , -3.36 )$&$ 2.2 _{ -0.7 }^{ +0.8 }$ $^{a}$&$ 5.86 $\\ % 2001ApJ...546..338P
4U~1254--690 &$( 303.48 , -6.42 )$&$ 13 \pm 3$ $^{b}$&$ 3.46 $\\ % 2003A&A...411L.487I
4U~1630--472 &$( 336.91 , 0.25 )$&$ 4 $ $^{c}$&$ 17.6 $\\ % 2002A&A...391..923G
4U~1636--53 &$( 332.92 , -4.82 )$&$ 6 \pm 0.5$ $^{d}$&$ 4.04 $\\ % 2006ApJ...639.1033G
4U~1702--429 &$( 343.89 , -1.32 )$&$ 6.2 \pm 0.9$ $^{e}$&$ 12.3 $\\ % 2004MNRAS.354..355J
4U~1705--44 &$( 343.32 , -2.34 )$&$ 8.4 \pm 1.2$ $^{e}$&$ 8.37 $\\ % 2004MNRAS.354..355J
4U~1728--16 &$( 8.51 , 9.04 )$&$ 4.4$ $^{c}$&$ 3.31 $\\ % 2002A&A...391..923G
4U~1728--34 &$( 354.30 , -0.15 )$&$ 5.3\pm 0.8$ $^{e}$&$ 13.9 $\\ % 2004MNRAS.354..355J
GX~9+9 &$( 8.51 , 9.04 )$&$ 4.4 $ $^{c}$&$ 3.31 $\\ % 2002A&A...391..923G
H1743--322 &$( 357.26 , -1.83 )$&$ 10.4 \pm 2.9$ $^{f}$&$ 8.31 $\\ % 2005ApJ...632..504C 
NGC~6624 &$( 2.79 , -7.91 )$&$ 7 $ $^{g}$&$ 2.33 $\\ % 2018MNRAS.478.1520B
EXO~1846--031 &$( 29.96 , -0.92 )$&$-$&$ 13.9 $\\ %  
GRS~1915+105 &$( 45.37 , -0.22 )$&$ 11 _{ -4 }^{ +1 }$ $^{e}$&$ 15.1 $\\ % 2004MNRAS.354..355J
GS~1826--238 &$( 9.27 , -6.09 )$&$ 7.5 \pm 0.5$ $^{h}$&$ 3.00 $\\ % 2000MNRAS.311..405K
GX~13+1 &$( 13.52 , 0.11 )$&$ 7\pm 1$ $^{i}$&$ 13.6 $\\ % 1999MNRAS.306..417B
GX~17+2 &$( 16.43 , 1.28 )$&$ 14 _{ -2.1 }^{ +2 }$ $^{e}$&$ 10.0 $\\ % 2004MNRAS.354..355J
GX~3+1 &$( 2.29 , 0.79 )$&$ 5 _{ -0.7 }^{ +0.8 }$ $^{j}$&$ 10.7 $\\ % 2001A&A...366..138O
GX~339--4 &$( 338.94 , -4.33 )$&$ 10 _{ -4 }^{ +5 }$ $^{k}$&$ 5.18 $\\ % 2004ApJ...609..317H
GX~340+0 &$( 339.59 , -0.08 )$&$ 11 $ $^{c}$&$ 20.0 $\\ % 2002A&A...391..923G
GX~349+2 &$( 349.10 , 2.75 )$&$ 9.2 $ $^{c}$&$ 6.13 $\\ % 2002A&A...391..923G
GX~354+0 &$( 354.30 , -0.15 )$&$ 5.3 \pm 0.8$ $^{e}$&$ 13.9 $\\ % 2004MNRAS.354..355J
GX~5--1 &$( 5.08 , -1.02 )$&$ 0.21 \pm 0.01$ $^{l}$&$ 10.4 $\\ % 2020yCat.1350....0G
V4641~Sgr &$( 6.77 , -4.79 )$&$ -$&$ 3.23 $\\ % 
X1543--62 &$( 321.76 , -6.34 )$&$ 7 $ $^{m}$&$ 3.79 $\\ % 2004ApJ...616L.139W
X1822--371 &$( 356.85 , -11.29 )$&$ 2.5 \pm 0.5$ $^{n}$&$ 1.40 $\\ % 1982ApJ...262..253M 
4U~1735--44 &$( 346.05 , -6.99 )$&$ 9.4 \pm 1.4$ $^{e}$&$ 3.96 $\\ % 2004MNRAS.354..355J
GX~9+1 &$( 9.08 , 1.15 )$&$ 4.4 \pm 1.3$ $^{o}$&$ 9.89 $\\ % 2005A&A...439..575I
4U~1916--053 &$( 31.36 , -8.46 )$&$ 8.8 \pm 1.3$ $^{e}$&$ 3.72 $\\ % 2004MNRAS.354..355J
4U~1957+11 &$( 51.31 , -9.33 )$&$ -$&$ 2.01 $\\ % 
A1744-361 &$( 354.12 , -4.19 )$&$ 9  $ $^{p}$&$ 4.44 $\\ % 2006ApJ...639L..31B
Cir~X--1&$( 322.12 , 0.04 )$&$ 9.2 _{ -1.4 }^{ +1.3 }$ $^{e}$&$ 16.4 $\\ % 2004MNRAS.354..355J
Cyg~X--2&$( 87.33 , -11.32 )$&$ 13.4 _{ -2 }^{ +1.9 }$ $^{e}$&$ 3.09 $\\ % 2004MNRAS.354..355J
Ser~X--1&$( 36.12 , 4.84 )$&$ 11.1 \pm 1.6$ $^{e}$&$ 5.42 $\\ % 2004MNRAS.354..355J
\hline
\multicolumn{4}{l}{ $N({\rm HI})$ in units of $10^{21}$cm$^{-2}$ }\\
\multicolumn{4}{l}{Distances obtained from $^a$\citet{pae01b};}\\
\multicolumn{4}{l}{$^b$\citet{int03};$^c$\citet{gri02}; }\\
\multicolumn{4}{l}{$^d$\citet{gal06};$^e$\citet{jon04};}\\
\multicolumn{4}{l}{$^f$\citet{cor05};$^g$\citet{bau18};}\\ 
\multicolumn{4}{l}{$^h$\citet{kon00};$^i$\citet{ban99};}\\ 
\multicolumn{4}{l}{$^j$\citet{oos01};$^k$\citet{hyn04};}\\ 
\multicolumn{4}{l}{$^l$\citet{gai20};$^m$\citet{wan04};}\\ 
\multicolumn{4}{l}{$^n$\citet{mas82};$^o$\citet{iar05};}\\ 
\multicolumn{4}{l}{$^p$\citet{bha06}.} 
\end{tabular} 
\end{table}
             \begin{figure*}
          \centering
\includegraphics[width=0.33\textwidth]{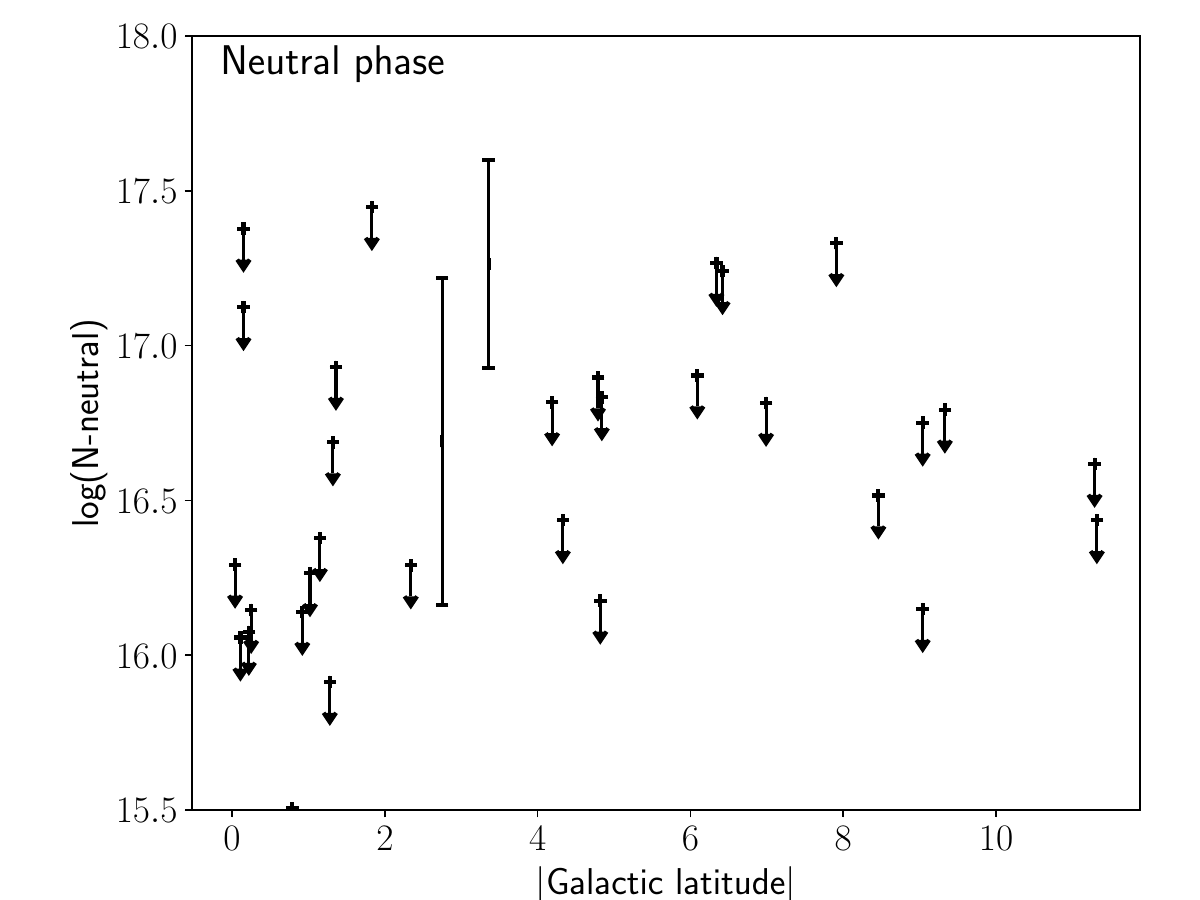} 
\includegraphics[width=0.33\textwidth]{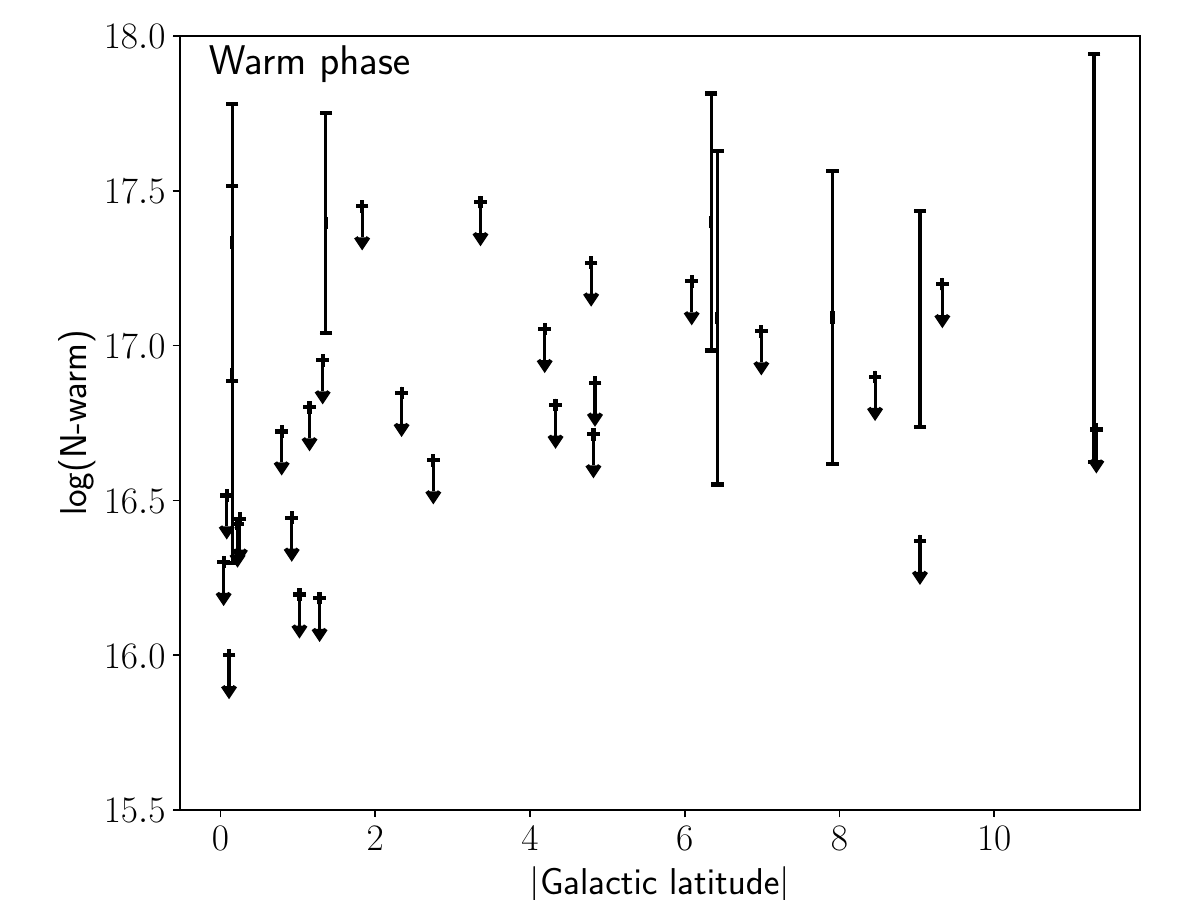} 
\includegraphics[width=0.33\textwidth]{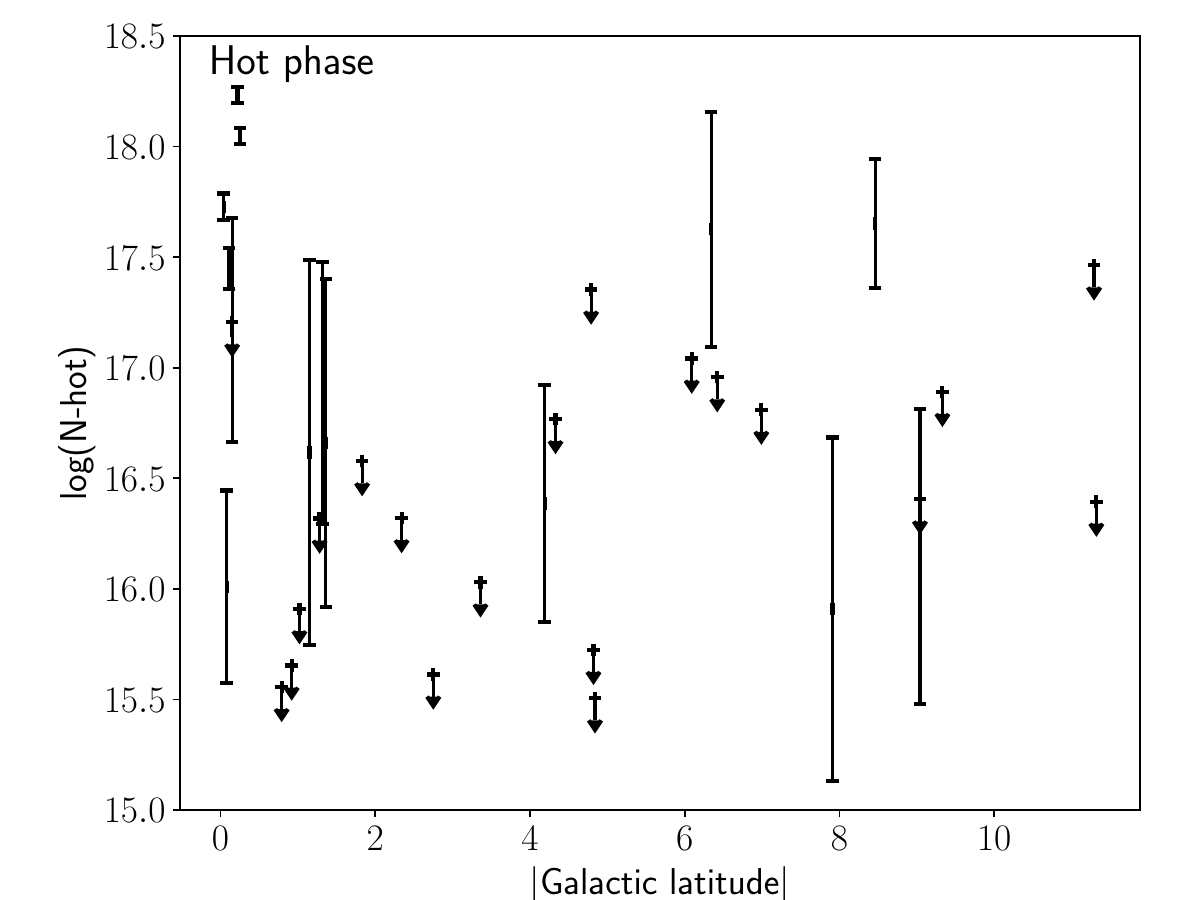} \\
\includegraphics[width=0.33\textwidth]{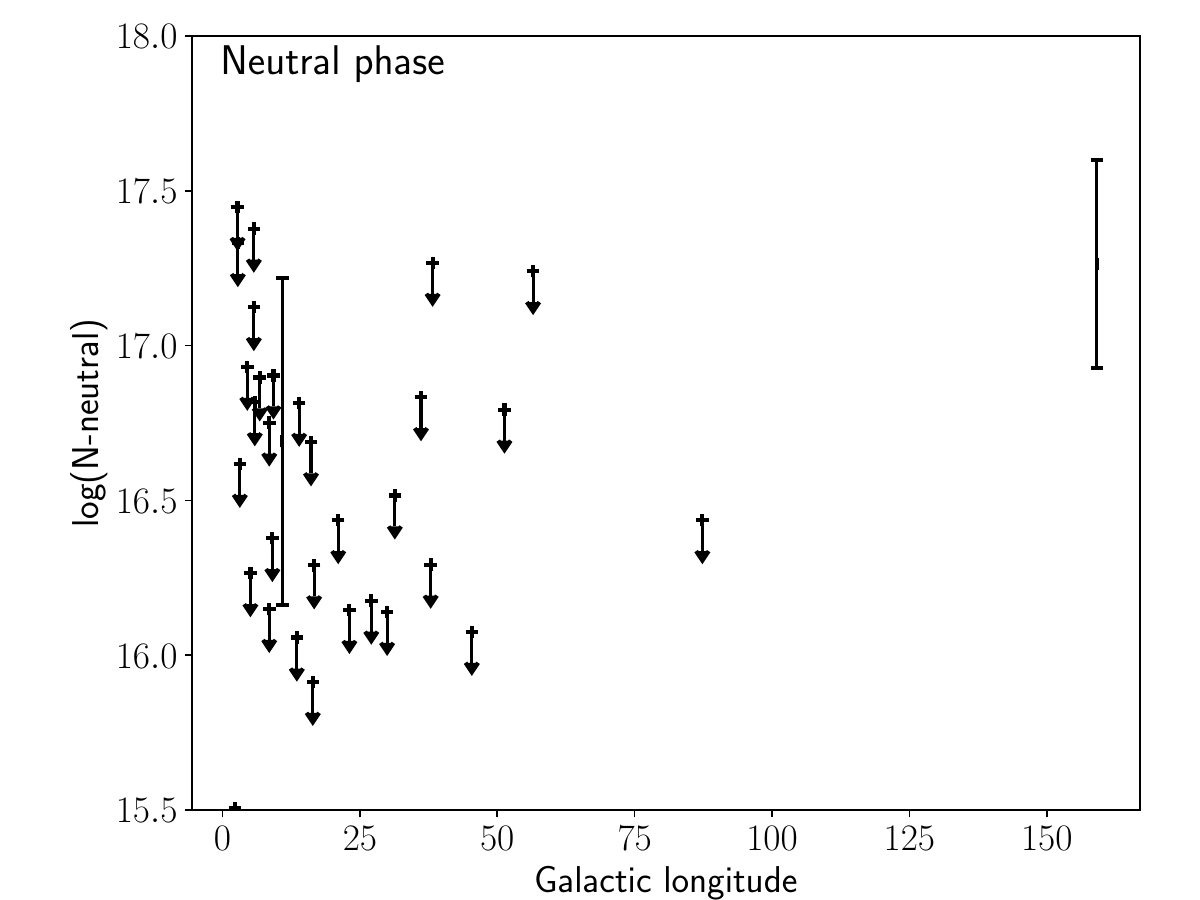} 
\includegraphics[width=0.33\textwidth]{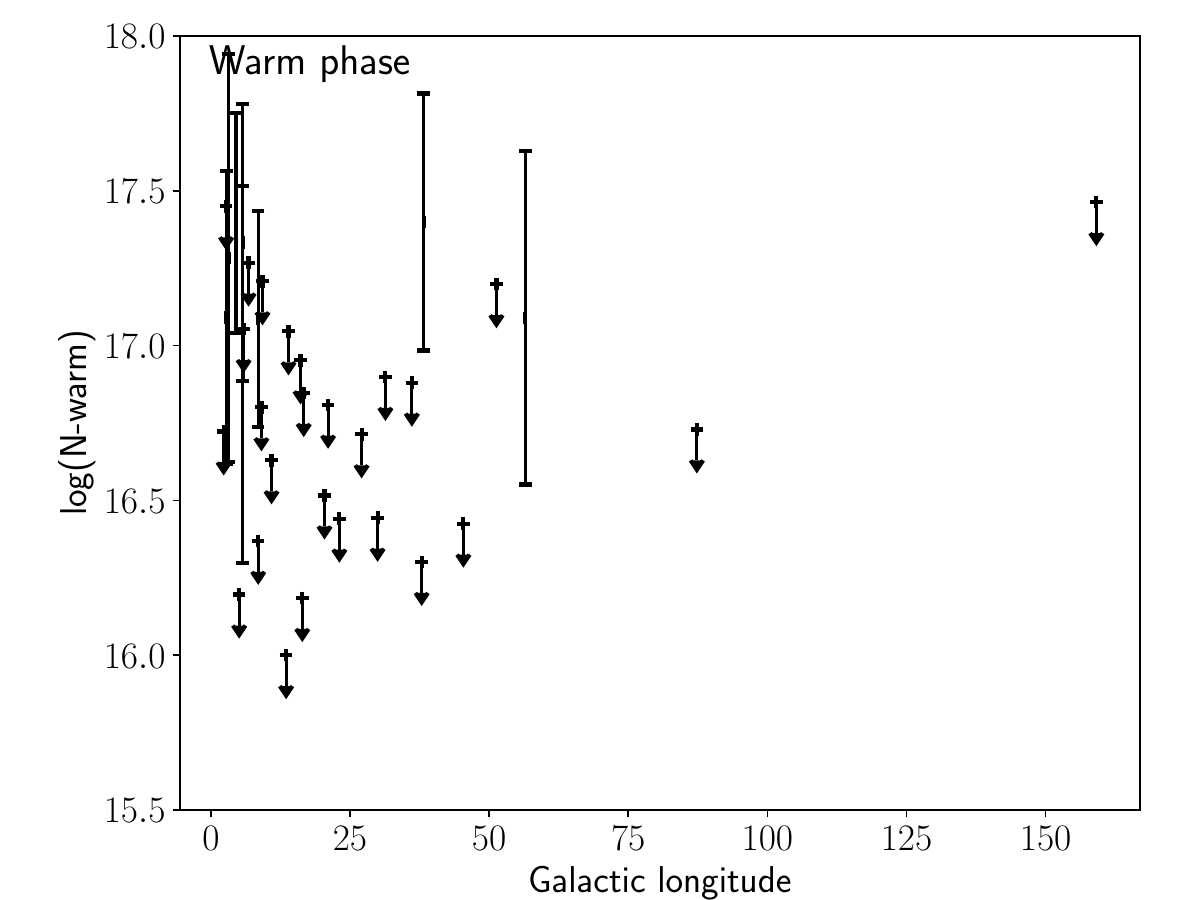} 
\includegraphics[width=0.33\textwidth]{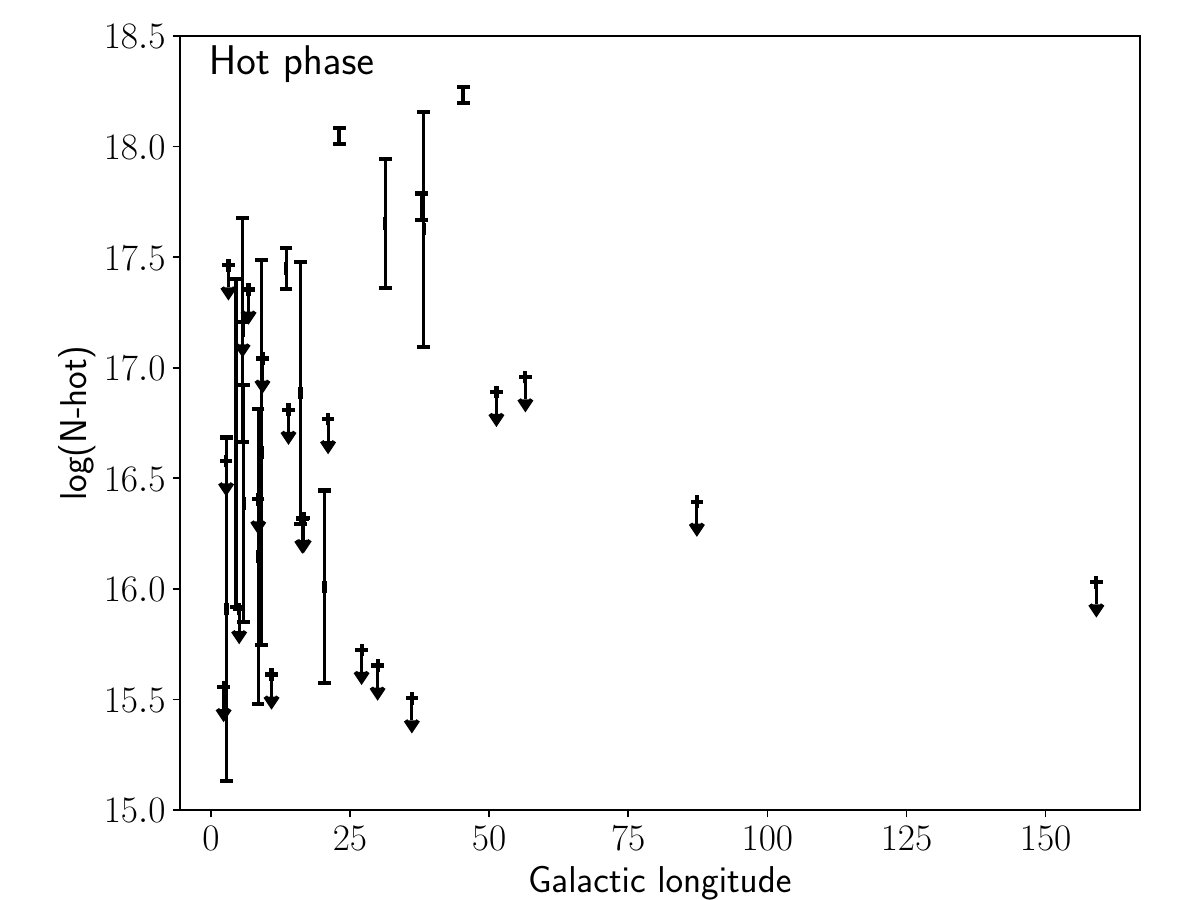} \\ 
      \caption{
      Ar column densities distribution for each ISM phase as function of Galactic latitude (top panels) and Galactic longitude (bottom panels). Upper values have not been included for illustrative purposes.
      }\label{fig_columns_latitude}
   \end{figure*}
             \begin{figure}
          \centering
\includegraphics[width=0.48\textwidth]{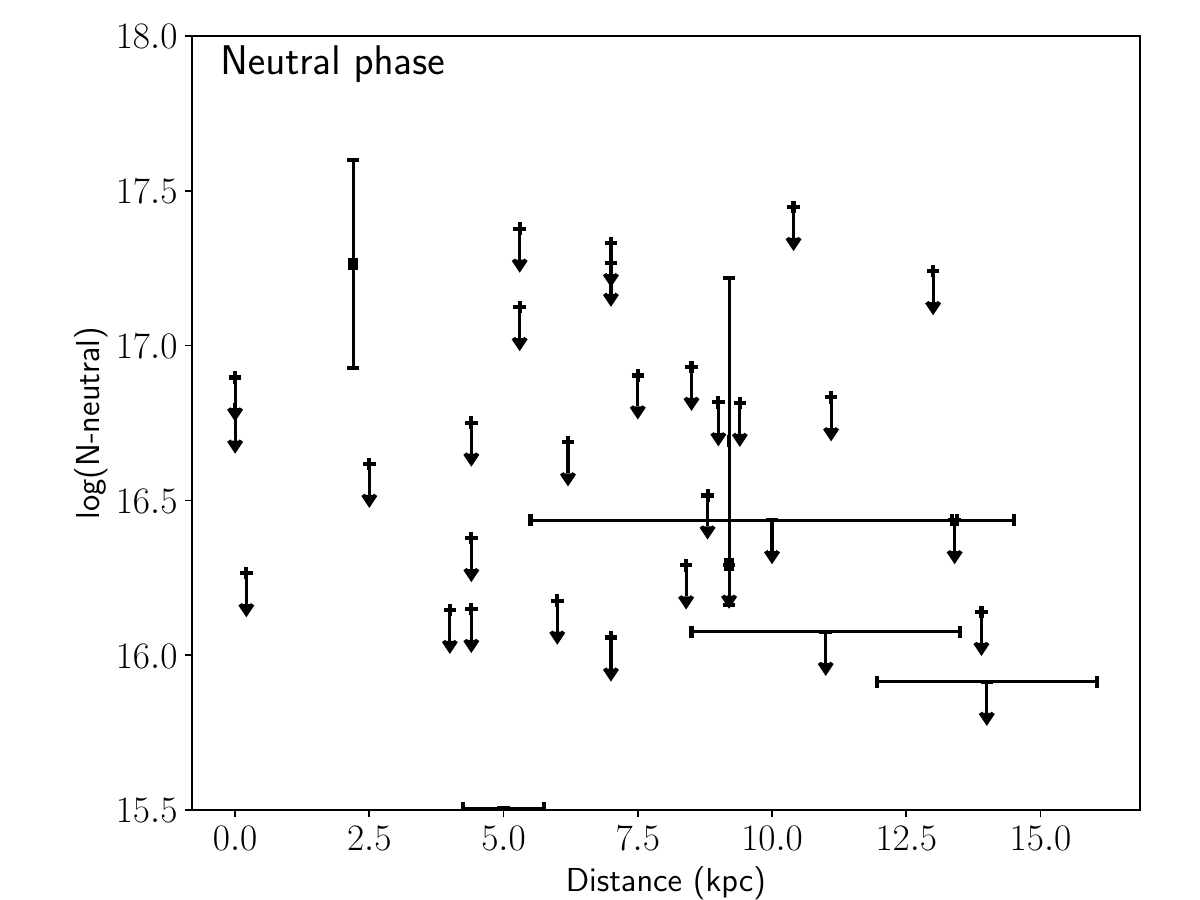} \\
\includegraphics[width=0.48\textwidth]{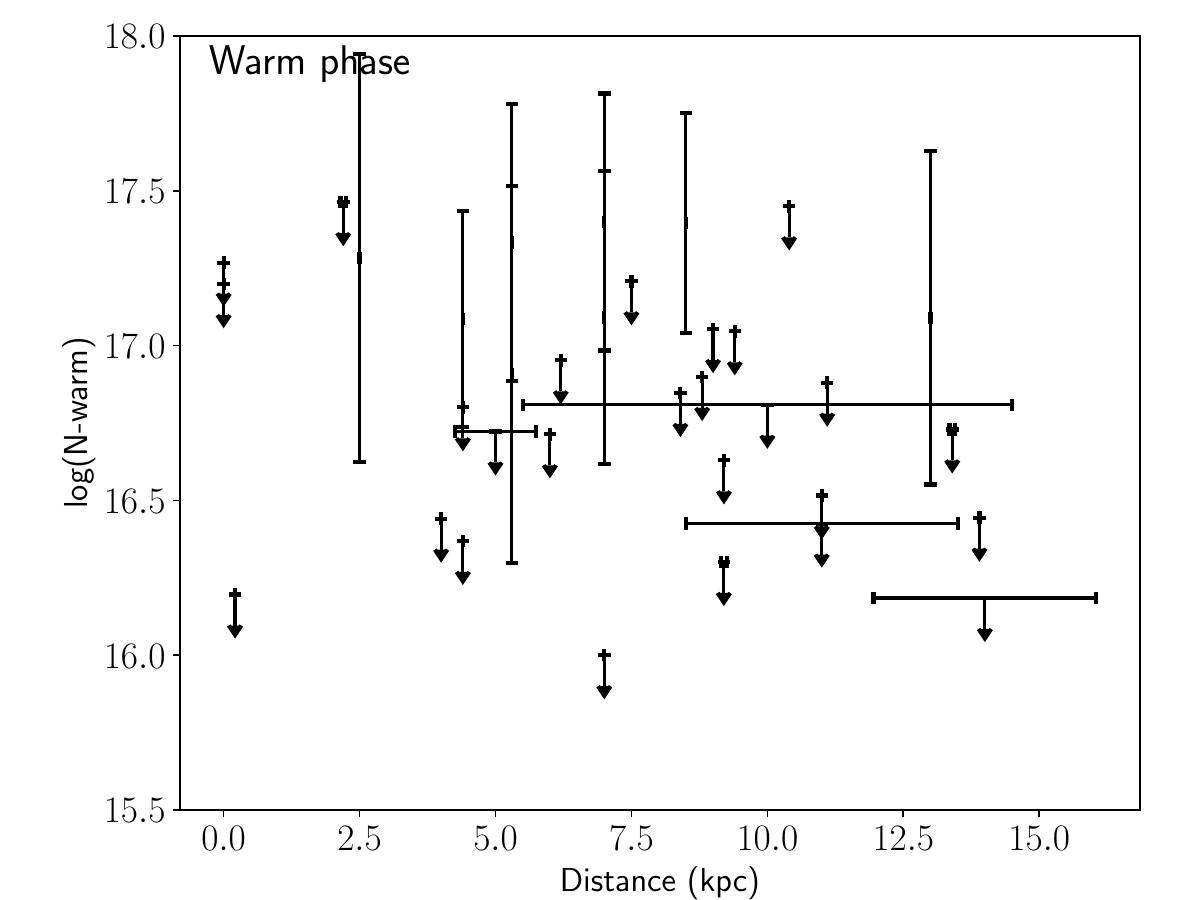} \\
\includegraphics[width=0.48\textwidth]{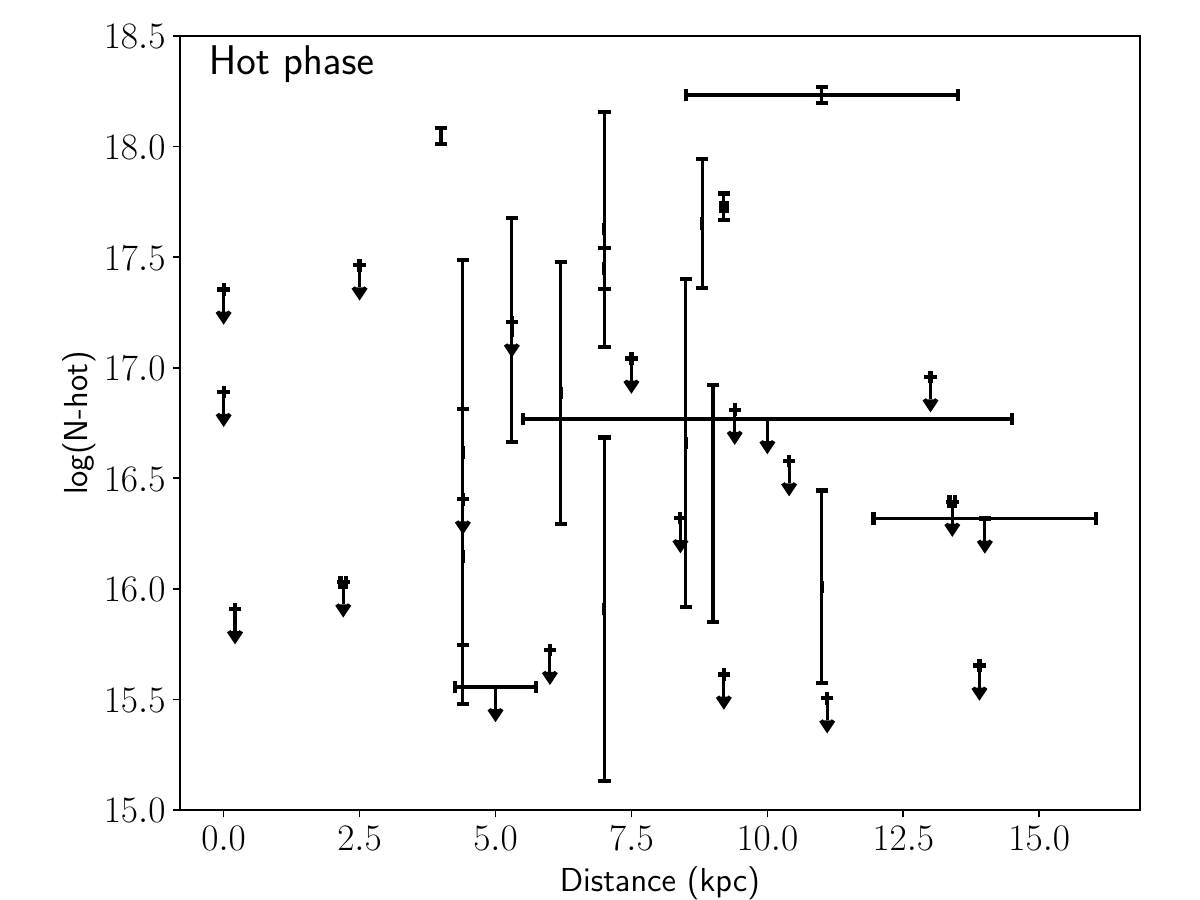} 
      \caption{
      Ar column densities distribution for each ISM phase as function of the distance. Upper values have not been included for illustrative purposes.
      }\label{fig_columns_distance}
   \end{figure}

\section{The Ar edge}\label{sec_dis}
 Table~\ref{tab_ismabs} shows the best-fit results.  
We distinguished between different phases of the gaseous ISM, categorizing them as cold ({\rm Ar}~{\sc i}), warm ({\rm Ar}~{\sc ii}+{\rm Ar}~{\sc iii}), hot ({\rm Ar}~{\sc xvi}+{\rm Ar}~{\sc xvii}+{\rm Ar}~{\sc xviii}).
These phases are defined following the analysis of the ISM X-ray absorption done by \citet{gat18a}, where cold component includes the neutral and molecular gas and corresponds to $T_{e}<1\times 10^{4}$~K, the warm component corresponds to $T_{e}\sim 5 \times 10^{4}$~K  and the hot component to $T_{e}\sim 1 \times 10^{6}$~K.
We noted that upper limits were obtained for relevant parameters for most sources. 
The obtained best-fit column densities are depicted in Figure~\ref{fig_data_density_fractions}, indicating column densities within similar ranges across different gaseous phases.

Figure~\ref{fig_columns_latitude} illustrates the distribution of column densities for each phase of the ISM concerning Galactic latitude (top panels) and Galactic longitude (bottom panels). 
Despite efforts, establishing a correlation with Galactic latitude remains challenging, with many sources yielding upper-limit results. 
This echoes findings by \citet{gat21,gat24}, who observed a consistent distribution of the warm-hot ISM component in their analysis of nitrogen and sulfur K-edge photoabsorption regions, alongside a decline in the cold component. 
Figure~\ref{fig_columns_distance} presents the column density distribution for each phase relative to the distance for applicable sources. 
Although there is a hint of a decreasing column density as a function of the distance for the cold-warm phases, discerning a clear correlation between these parameters proves elusive.

While this study represents the first in-depth exploration of Argon X-ray absorption using high-resolution spectra, it is pertinent to consider comparisons with previous research. 
Prior analyses of the ISM utilizing X-ray absorption have revealed a predominance of the neutral component, with mass fractions for various phases in the Galactic disc approximately at $\sim~90\%$ for neutral, $\sim~8\%$ for warm, and $\sim~2\%$ for hot phases \citep[e.g.,][]{yao06,pin13,gat18a}. 
However, due to predominantly obtaining upper limits in our study, accurate computation of mass fractions for all sources proves challenging. 
We do not consider ionization equilibrium for argon ionic species, as the column densities in the {\it ISMabs} model are treated as free parameters. 
Therefore, the temperature of the hot phase may not be sufficiently high to yield highly ionized Ar. 
Furthermore, the hot phase may include contributions from ionized static absorbers intrinsic to the source \citep[see, for example,][]{gat20b}. 
 It is commonly assumed that the neutral component of the ISM exponentially decreases along the perpendicular direction to the Galactic plane, with larger column densities observed near the Galactic center \citep[see, e.g.,][]{rob03,kalb09,gat24b}. 
Nevertheless, argon depletion into dust may deviate from such distribution patterns, which could impact the observed column densities in both the cold and warm ISM atomic phases.
Depending on the level of depletion, the X-ray absorption lines attributable to gas-phase argon would be weaker than expected, thus potentially leading to an underestimation of the argon abundance and also affecting the benchmarking of the atomic data \citep{cos22}.  
While a comprehensive thermodynamic analysis of the ISM component, incorporating dust depletion effects, is beyond the scope of this study, acknowledging its potential impact on the observed column densities highlights an important area for future research.

  \begin{table*}
\caption{
\label{tab_ismabs}
Best-fit argon column densities obtained.
 }
%\scriptsize
\centering
\begin{tabular}{lccccccc}
\hline  
Source  & Ar\,{\sc i} &  Ar\,{\sc ii}  &  Ar\,{\sc iii} & Ar\,{\sc xvi} &  Ar\,{\sc xvii}  &  Ar\,{\sc xviii}& cstat/d.o.f.  \\
 \hline
\hline  
4U1254-690& $<17.4$& $<11.5$& $12.3^{+15.6}_{-9.0}$& $<1.4$& $<7.2$& $<0.5$& $561.7/471$\\  
4U1630-472& $<1.4$& $<1.4$& $<1.3$& $108.6^{+9.3}_{-6.3}$& $<0.2$& $3.1^{+1.7}_{-1.3}$& $657.4/471$\\  
4U1636-53& $<1.5$& $<2.6$& $<2.6$& $<0.4$& $<0.1$& $<0.0$& $696.9/471$\\  
4U1702-429& $<4.9$& $<4.8$& $<4.1$& $7.7^{+21.2}_{-5.0}$& $<0.2$& $<11.9$& $566.1/471$\\  
4U1705-44& $<2.0$& $<2.7$& $<4.3$& $<0.4$& $<1.4$& $<0.3$ &$668.1/471$\\  
4U1728-16& $<5.6$& $<8.2$& $12.2^{+10.0}_{-6.1}$& $1.4^{+4.4}_{-1.0}$& $<0.5$& $<2.9$& $539.0/471$\\  
4U1728-34& $<13.3$& $<48.6$& $21.1^{+16.8}_{-10.0}$& $<10.1$& $<2.1$& $<3.8$& $423.1/471$\\  
GX9+9& $<1.4$& $<1.2$& $<1.1$& $<0.8$& $<0.8$& $<1.0$& $514.4/471$\\  
H1743-322& $<28.0$& $<20.0$& $<8.1$& $<2.1$& $<0.3$& $<1.4$ &$460.7/471$\\  
NGC6624& $<21.4$& $12.3^{+11.7}_{-8.0}$& $<14.4$& $<3.0$& $0.8^{+2.5}_{-0.6}$& $<0.8$ &$479.8/471$\\  
EXO1846-031& $<1.4$& $<1.4$& $<1.4$& $<0.3$& $<0.1$& $<0.1$& $629.2/471$\\  
GRS1758-258& $<8.5$& $<30.4$& $24.9^{+20.9}_{-10.0}$& $<11.4$& $<3.1$& $4.6^{+27.9}_{-3.0}$& $596.2/471$\\  
GRS1915+105& $<1.2$& $<1.3$& $<1.4$& $113.9^{+6.8}_{-6.7}$& $4.3^{+1.5}_{-1.2}$& $52.8^{+5.9}_{-5.5}$& $462.1/471$\\  
GS1826-238& $<8.0$& $<8.7$& $<7.4$& $<2.4$& $<3.7$& $<4.9$ &$391.6/471$\\  
GX13+1& $<1.1$& $<0.5$& $<0.5$& $24.5^{+5.5}_{-4.1}$& $<0.2$& $3.6^{+1.3}_{-1.1}$& $626.1/471$\\  
GX17+2& $<0.8$& $<0.8$& $<0.7$& $<1.0$& $<0.9$& $<0.2$& $516.0/471$\\  
GX3+1& $<0.3$& $<3.0$& $<2.3$& $<0.2$& $<0.2$& $<0.1$ &$589.7/471$\\  
GX339-4& $<2.7$& $<3.0$& $<3.4$& $<2.6$& $<1.2$& $<2.1$ &$487.5/471$\\  
GX340+0& $<0.2$& $<2.1$& $<1.2$& $<0.3$& $<0.1$& $1.0^{+1.2}_{-0.7}$& $646.5/471$\\  
GX349+2& $4.9^{+5.4}_{-4.0}$& $<2.5$& $<1.7$& $<0.1$& $<0.2$& $<0.1$& $556.4/471$\\  
GX354+0& $<23.9$& $<10.9$& $8.1^{+15.2}_{-6.0}$& $14.8^{+15.7}_{-11.6}$& $<1.4$& $<0.4$ &$458.7/471$\\ 
GX5-1& $<1.8$& $<0.8$& $<0.8$& $<0.7$& $<0.0$& $<0.1$& $644.6/471$\\  
V4641Sgr& $<7.9$& $<9.1$& $<9.4$& $<17.5$& $<4.7$& $<0.4$ &$454.7/471$\\ 
X1543-62& $<18.5$& $<41.3$& $22.9^{+22.9}_{-10.0}$& $42.3^{+43.2}_{-33.7}$& $<11.3$& $<2.0$& $476.9/471$\\  
X1822-371& $<4.1$& $<20.2$& $19.2^{+15.4}_{-16.6}$& $<23.5$& $<5.1$& $<0.5$ &$545.4/471$\\  
4U1735-44& $<6.5$& $<5.9$& $<5.2$& $<5.6$& $<0.6$& $<0.4$ &$449.2/471$\\  
GX9+1& $<2.4$& $<2.9$& $<3.4$& $4.1^{+15.5}_{-3.7}$& $<1.4$& $<0.9$ &$576.6/471$\\  
4U1916-053& $<3.3$& $<3.7$& $<4.2$& $29.6^{+23.7}_{-12.0}$& $4.1^{+6.2}_{-3.0}$& $11.3^{+11.9}_{-7.4}$& $530.6/471$\\  
4U1957+11& $<6.2$& $<8.5$& $<7.3$& $<2.3$& $<0.6$& $<4.8$& $508.2/471$\\  
A1744-361& $<6.6$& $<5.9$& $<5.4$& $<0.7$& $2.4^{+3.2}_{-1.7}$& $<2.1$& $455.6/471$\\  
CIRX-1& $<2.0$& $<1.0$& $<1.0$& $49.1^{+7.7}_{-4.7}$& $4.4^{+1.4}_{-1.2}$& $<0.0$& $607.3/471$\\  
CYGX-2& $<2.7$& $<2.5$& $<2.9$& $<1.0$& $<1.1$& $<0.3$& $464.3/471$\\  
SERX-1& $<6.8$& $<5.8$& $<1.8$& $<0.1$& $<0.0$& $<0.2$& $442.3/471$\\  
\hline 
\multicolumn{7}{l}{ Column densities in units of $10^{16}$cm$^{-2}$.}
 \end{tabular}
\end{table*}

 \subsection{Future prospects}\label{sec_sim}
\begin{figure}
          \centering
\includegraphics[width=0.48\textwidth]{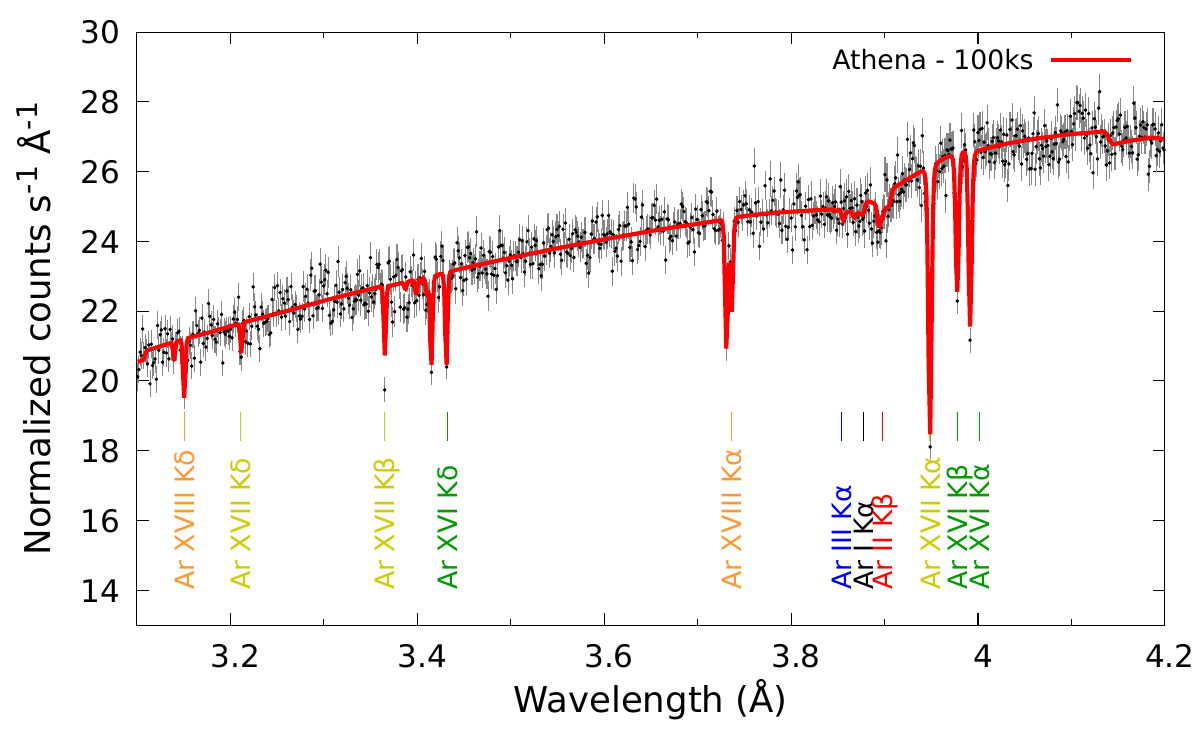} \\
\includegraphics[width=0.48\textwidth]{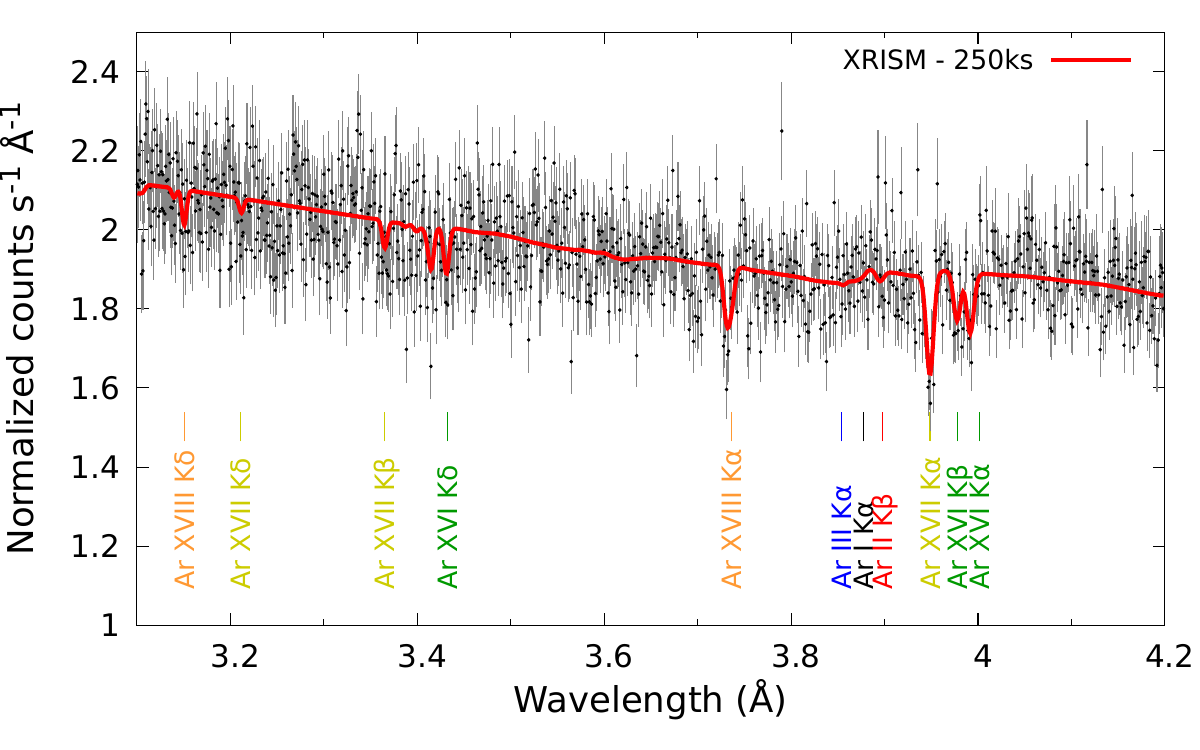}
      \caption{
      {\it Athena} X-IFU (top panel) and {\it XRISM} (bottom panel) simulations of the Ar K-edge photoabsorption region for a Galactic source (e.g. 4U~1916-053). The total exposure time is indicated. 
      }\label{ath_sim}
   \end{figure}

Future advancements in X-ray observatories will allow us to resolve the intricate $K\alpha$ resonance lines across various argon ionic species. 
An illustration of this potential is depicted in Figure~\ref{ath_sim} (top panel), showing a simulation focused on a Galactic source, specifically 4U~1916-053, achieved through {\it Athena} \citep{nan13}. 
This simulation, done with the {\tt sixte} software \citep{dau19}, integrates the response files of the Athena X-ray Integral Unit (X-IFU), distributed after the reformulation of the Athena mission \citep{bar23}. 
In particular, we adopt an instrumental spectral resolution of 3 eV and a nominal X-IFU configuration without a filter.
 We also simulated a 250~ks {\it XRISM} observation using the same model in combination with the response files available for the {\it XRISM} Guest Observer Cycle 1 program (Figure~\ref{ath_sim}, bottom panel).
The {\it XRISM} observatory was launched successfully on September 7, releasing its first light on January 05 and showing unprecedented high-resolution spectra for the Perseus cluster and the supernova remnant N132D, becoming the current X-ray community observatory.

The plot shows the remarkable capabilities of the instrument, where prominent resonance absorption lines emerge distinctly, facilitating comprehensive investigations into the multiphase ISM. 
Furthermore, the simultaneous measurement of K$\alpha$ and K$\beta$ absorption lines for identical ions promises more accurate constraints on abundances, broadening effects, and ionization state estimatiWe note that this simulation exclusively accounts for the gaseous component of the ISM. 
Although the cumulative dust contribution could be quantified, distinguishing between various dust samples poses a more intricate challenge. 
For a comprehensive understanding, \citet{cos19} conducted an exhaustive dust simulation tailored for {\it Athena}.

\section{Conclusions}\label{sec_con}
We have analyzed the argon K-edge X-ray absorption region (3.1-4.2 \AA) using {\it Chandra} high-resolution spectra of 33 LMXBs. 
 Each source was fitted with a simple {\tt powerlaw} for the continuum and a modified version of the {\it ISMabs} model.
Our study provides detailed calculations of new photoabsorption cross-sections for {\rm Ar}~{\sc i}, {\rm Ar}~{\sc ii} and {\rm Ar}~{\sc xvi}. 
 Using this model, we derived column density estimates for {\rm Ar}~{\sc i}, {\rm Ar}~{\sc ii}, {\rm Ar}~{\sc iii}, {\rm Ar}~{\sc xvi}, {\rm Ar}~{\sc xvii} and {\rm Ar}~{\sc xviii} ionic species 
Our analysis revealed that individual K$\alpha$ doublets/triplets could not be resolved, leading to upper limits for most sources. 
Furthermore, we observed no correlation between Galactic longitude and latitude, although there were indications of decreasing column density with distance. 
Finally, our results from the {\it Athena} X-IFU and {\it XRISM} simulations demonstrate that these observatories will enable unprecedented precision in atomic data benchmarking.
Their high-resolution capabilities will allow us to resolve intricate spectral features that were previously inaccessible, leading to more accurate atomic models and a deeper understanding of the physical conditions in various astrophysical environments.

\subsection*{Data availability}
Observations analyzed in this article are available in the Chandra Grating-Data Archive and Catalog (TGCat)  (\url{http://tgcat.mit.edu/about.php}). 
The {\tt ISMabs} model is included in the {\sc xspec} data analysis software (\url{https://heasarc.gsfc.nasa.gov/xanadu/xspec/}). 
This research was carried out on the High Performance Computing resources of the cobra cluster at the Max Planck Comput-ing and Data Facility (MPCDF) in Garching operated by the Max Planck Society (MPG)

%\begin{acknowledgements} 
% 
%\end{acknowledgements}

\bibliographystyle{aa}

\begin{appendix}
\section{{\it Chandra} Observation IDs}\label{sec_app}

Table~\ref{tab_obsids} list the {\it Chandra} observation IDs analyzed in this work

\begin{table*}
\caption{\label{tab_obsids}List of {\it Chandra} observations IDs analyzed.}
%\tiny
\centering
\begin{tabular}{lc}
\hline
Source   & ObsIDs\\ 
\hline  
4U~0614+091 & 10759,10760,10857,10858  \\
4U~1254--690 & 3823  \\
4U~1630--472 & 13714,13715,13716,13717,15511,15524,22376,22377, 2237  \\
4U~1636--53 & 105,1939,20791,21099,21100,22701,22936,24625,24626  \\
4U~1702--429 & 11045  \\
4U~1705--44 & 18086,1923,19451,20082,5500  \\
4U~1728--16 & 703  \\
4U~1728--34 & 2748 \\
GX~9+9 & 11072 \\
H1743--322 & 16738,16739,16740,16741,17679,17680,8985,8986  \\
NGC~6624 & 1021,1022  \\
EXO~1846--031 & 20899,20900,21237,21237,21238  \\
GRS~1915+105 & 22213,22885,22886,23435,24663  \\
GS~1826--238 & 2739  \\
GX~13+1 & 11814,11815,11816,11817,20191,20192,20193,20194,2708  \\
GX~17+2 & 11088  \\
GX~3+1 & 16307,16492,18615,19890,19907,19957,19958,27271,27272,27273  \\
GX~339--4 & 4420,5475,6290  \\
GX~340+0 & 18085,1921,19450,20099  \\
GX~349+2 & 13220,14256,18084,3354,715  \\
GX~354+0 & 19452,20106,20107  \\
GX~5--1 & 19449,20119,22411,22412,22413,22414,22415,22416,716  \\
V4641~Sgr & 22389,23158  \\
X1543--62 & 702  \\
X1822--371 & 9076,9858  \\
4U~1735--44 & 704  \\
GX~9+1 & 717,717  \\
4U~1916--053 & 20171,20172,21103,21105,21106,21662,21663,21666,4584  \\
4U~1957+11 & 10659,10661,4552  \\
A1744-361 & 9042,9884,9885  \\
Cir~X--1 & 12235,1700,18990,1905,1906,1907,20093,20094,5478,6148,706,8993  \\
Cyg~X--2 & 1016,1102  \\
Ser~X--1 & 17485,17600,700 \\
\hline 
\end{tabular} 
\end{table*}

\end{appendix}

\end{document}